\documentclass{LT23auth}
\usepackage{graphicx}

\begin{document}

\begin{frontmatter}

\title{Anisotropic magnetic behavior of GdBa$_{2}$Cu$_{3}$O$_{6+y}$
single crystals}

\author[address2]{Vladimir N. Narozhnyi$^{a,}$\thanksref{thank1}},
\author[address1]{Dieter Eckert},
\author[address1]{G\"unter Fuchs},
\author[address3]{Vladimir Nekvasil},
\author[address1]{Karl-Hartmut M\"uller}

\address[address1]{Leibniz-Institut f\"ur Festk\"orper- und Werkstoffforschung Dresden,
PO Box 270116, D-01171 Dresden, Germany}

\address[address2]{Institute for High
Pressure Physics, Russian Academy of Sciences, Troitsk, Moscow Region,
142190, Russia}

\address[address3]{Institute of Physics, Czech Academy of
Sciences, Cukrovarnick\'a 10, 16253 Praha 6, Czech Republic}

\thanks[thank1]{Corresponding author. Present address: Leibniz-Institut f\"ur
Festk\"orper- und Werkstoffforschung Dresden, PO Box 270116, D-01171
Dresden, Germany. E-mail: narozh@ifw-dresden.de}

\begin{abstract}
Magnetic properties of high-quality Al-free nonsuperconducting $\rm
GdBa_2Cu_3O_{6+y}$ single crystals grown by flux method have been
studied. The magnetic anisotropy below the N\'eel temperature
$T_\mathrm{N}\approx 2.3~$K corresponds to the direction of $\rm
Gd^{3+}$ magnetic moments along the tetragonal $c$-axis. At
$T<T_\mathrm{N}$ clear indications of spin-flop transitions for
$H\parallel c$ have been observed on magnetization curves at
$H_\mathrm{sf}\approx 10~$kOe. Magnetic phase diagrams have been
obtained for $H\parallel c$ as well as for $H\perp c$. A pronounced
anisotropy in the magnetic susceptibility (unexpected for Gd-based
compounds) has been found above $T_\mathrm{N}$.
\end{abstract}

\end{frontmatter}

Collinear antiferromagnetic ordering was found for $\rm
GdBa_2Cu_3O_{6+y}$ (Gd1236) with the Gd magnetic moments directed along
the $c$-axis \cite{Mook_PRB88,Meyer_JPhysF87}. So far magnetism in
Gd1236 single crystals were studied on Al-containing samples
\cite{Djakonov_PhC91} or by indirect methods (e.g., NMR
\cite{Nehrke_PRB95}). However, the reported results are controversial.

In this work we report on the magnetic properties of high quality
Al-free nonsuperconducting Gd1236 single crystals grown in Pt crucibles
by the flux method \cite{Nar_JMMM96}. Atomic absorption spectroscopy
has shown that the Pt contamination does not exceed
$3\cdot10^{-3}$~at.~\% \cite{Nar_PhB00}. To reduce the oxygen
concentration the samples were annealed at $T=$~600~C under high vacuum
during 4~days. The magnetization $M$ was measured by SQUID
magnetometer. The data are compared with the results obtained earlier
for $\rm PrBa_2Cu_3O_{7-y}$ (Pr123) \cite{Nar_PhC99} as well as for the
Gd-sublattice of mixed $\rm Gd_{1-x}Pr_xBa_2Cu_3O_{7-y}$ [(Gd-Pr)123]
single crystals \cite{Nar_PhB00}.

\begin{figure}[btp]
\begin{center}\leavevmode
\includegraphics[width=0.8\linewidth]{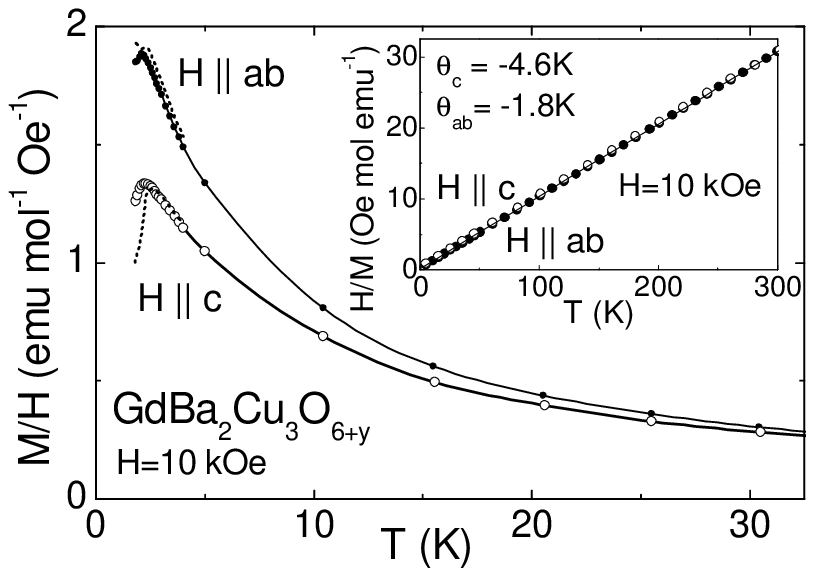}
\caption{ Temperature dependence of the magnetic susceptibility
(determined as $M/H$ at $H=10$~kOe) of a GdBa$_{2}$Cu$_{3}$O$_{6+y}$
single crystal for $H\parallel c$ (open circles) and $H\perp c$ (full
circles). Dotted lines - the data for $H=1$~kOe. Inset: temperature
dependence of inverse magnetic susceptibility $H/M$. The line shows the
best fit to the Curie-Weiss law for $H\perp c$.}
\label{fig_m_t}\end{center}\end{figure}

The temperature dependence of susceptibility $\chi=M/H$, shown in Fig.\
\ref{fig_m_t} for two directions of the magnetic field $H$, has a
maximum at $T\approx T_\mathrm{N}$. The position of the maximum is
field dependent. The anomaly is more pronounced for $H\parallel c$ (at
least for small fields), which corresponds to the $c$-axis direction of
the $\rm Gd^{3+}$ magnetic moments (in accord with neutron diffraction
and M\"{o}ssbauer spectrometry \cite{Mook_PRB88,Meyer_JPhysF87}). The
maximum in $\chi(T)$ disappears for $H\geq 15$~kOe at $T\geq 1.8$~K.

The $\chi^{-1}(T)$ curves can be well fitted to the Curie-Weiss law
$\chi^{-1}(T)=(T-\theta)/C$ for $5\leq T\leq 300$~K. The values of
$\theta$ and the effective magnetic moment $m_\mathrm{eff}$ (determined
from the Curie constants $C$) are -4.6~K and 8.90~$m_\mathrm{B}$ and
-1.8~K and 8.88~$m_\mathrm{B}$ for $H\parallel c$ and $H \perp c$,
respectively. $m_\mathrm{eff}^{c}$  and $m_\mathrm{eff}^{ab}$ are very
close to each other as expected for the spin-only magnetic moment of
Gd. The values of $m_\mathrm{eff}$ are little higher than the value of
the free $\rm Gd^{3+}$ ion. This difference may be connected with some
error in determination of the small mass of the sample ($m=0.50$~mg;
the accuracy of mass determination in this case is about $10\%$).

The anisotropy in $\chi$ in paramagnetic state is clearly seen in this
figure. Phenomenologically the observed anisotropy can be connected
with the difference between $\theta_\mathrm{c}$ and
$\theta_\mathrm{ab}$. (From the fit the accuracy in $\theta$
determination is better than 0.1 K.) It is found that
$|\theta_\mathrm{c}| > |\theta_\mathrm{ab}|$. The sign of magnetic
anisotropy for Gd1236 is the opposite to the observed for Pr123
\cite{Nar_PhB00,Nar_PhC99}. The different signs of magnetic anisotropy
for Gd- and Pr-sublattices explain the crossover of magnetic anisotropy
reported for (Gd-Pr)123 single crystals \cite{Nar_PhB00}.

\begin{figure}[btp]
\begin{center}\leavevmode
\includegraphics[width=0.8\linewidth]{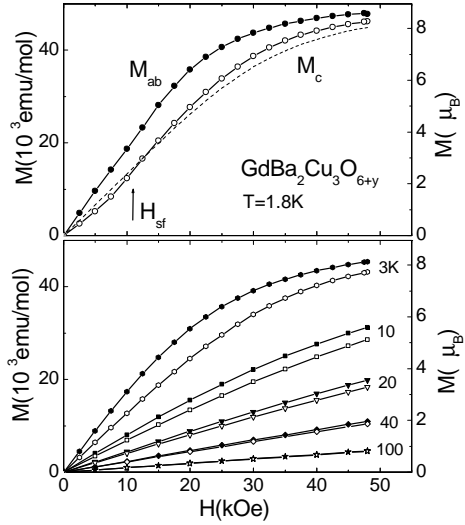}
\caption{ Magnetization of the GdBa$_{2}$Cu$_{3}$O$_{6+y}$ single
crystal for $H\parallel c$ (open symbols) and $H\perp c$ (full
symbols). Full lines are guides for eye. The dashed line shows the data
for $H\parallel c$ at $T=2.4$~K which is slightly above $T_\mathrm{N}$.
} \label{fig_m_h}\end{center}\end{figure}

$M(H)$ curves clearly show an anisotropy below as well as above
$T_\mathrm{N}$, see Fig.\ \ref{fig_m_h}. At lowest $T$ there is a clear
tendency for isotropization in high fields. $M_\mathrm{ab} >
M_\mathrm{c}$ is in accord with our data obtained earlier from
subtraction of (Y-Pr)123 data from (Gd-Pr)123 results \cite{Nar_PhB00}.
The general agreement with these previously indirectly obtained results
on the magnetization of the Gd-sublattice is fair. Even the values and
the anisotropy of $\theta$ correspond well to the directly measured for
Gd1236.

\begin{figure}[btp]
\begin{center}\leavevmode
\includegraphics[width=0.795\linewidth]{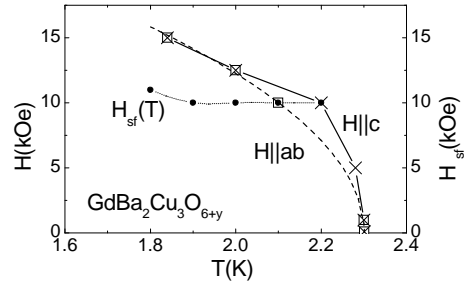}
\caption{ Magnetic phase diagram of GdBa$_{2}$Cu$_{3}$O$_{6+y}$ for
$H\parallel c$ (crosses and full line) and $H\perp c$ (open squares and
dashed line). The temperature dependence of the spin-flop field is
shown by the dotted line and full circles.}
\label{fig_ph_dia}\end{center}\end{figure}

Below $T_\mathrm{N}$, a distinct indication of a spin-flop transition
can be seen for $H\parallel c$, see Fig.\ \ref{fig_m_h}. The spin-flop
field $H_\mathrm{sf}$ was determined as an inflection point of the
$M(H)$ dependence at $T<T_\mathrm{N}$. $H_\mathrm{sf}$ is practically
temperature independent in the studied temperature interval, see Fig.\
\ref{fig_ph_dia}. As expected, the spin-flop transition disappears
above $T_\mathrm{N}$. No anomaly has been detected for $H\perp c$ at
all $T$.

A magnetic phase diagram is constructed from the field dependence of
$T_\mathrm{N}$, determined for two directions of $H$ (see Fig.\
\ref{fig_ph_dia}). The field dependence of $T_\mathrm{N}$ for $H\perp
c$ is described by a quadratic dependence similar to that observed by
us earlier for Pr-1237 \cite{Nar_PhC99}. Below $H_\mathrm{sf}$ the
$T_\mathrm{N}(H)$ dependence for $H\parallel c$ is weaker than for
$H\perp c$. At the same time above $H_\mathrm{sf}$ the
$T_\mathrm{N}(H)$ dependencies are close for both directions of $H$.

The pronounced magnetic anisotropy found for Gd1236 may be connected
with several mechanisms including: (i) dipole-dipole interaction of the
Gd ions; (ii) interaction between Gd and Cu sublattices; (iii)
anisotropic exchange interaction; (iv) crystal-field effects on the
excited $\rm Gd^{3+}$ states. Further investigations are necessary to
clarify the situation.

\begin{ack}
This work was supported by DFG (grant MU1015/4-2), RFBR (grant
01-02-04002) and GA CR (grant 202/00/1602).
\end{ack}

%
%

\end{document}